\documentclass[11pt]{amsart}

\usepackage{geometry}                
\usepackage{graphicx}
\usepackage{amssymb}
\usepackage{epstopdf}
\usepackage[sorting=ynt, 
citestyle=authoryear
]{biblatex}
\bibliography{ref} %

\title{A New Taxonomy for Symbiotic EM Sensors}
\author{MR Inggs}
\address{Dept Electrical Engineering, University of Cape Town, Rondebosch 7701, South Africa}

\email{Michael.Inggs@uct.ac.za}
\author{AK Mishra}
\address{Dept Electrical Engineering, University of Cape Town, Rondebosch 7701, South Africa}

\email{Amit.Mishra@uct.ac.za}

\begin{document}
\maketitle

\section{The Reality of Spectrum Congestion}
It is clear that the EM spectrum is now rapidly reaching saturation, especially for frequencies below 10~GHz~\footnote{$www.ntia.doc.gov/files/ntia/publications/january\_2016\_spectrum\_wall\_chart.pdf$}. Governments, who influence the regulatory authorities around the world, have resorted to auctioning the use of spectrum, in a sense to gauge the importance of a particular user. Billions of USD are being paid for modest bandwidths.

The earth observation, astronomy and similar science driven communities cannot compete financially with such a pressure system, so this is where governments have to step in and assess / regulate the situation.

It has been a pleasure to see a situation where the communications and broadcast communities have come together to formulate sharing of an important part of the spectrum (roughly, 50~MHz to 800~MHz) in an IEEE standard, IEEE802.22. This standard (known as the ``TV White Space Network''  (built on lower level standards) shows a way that fixed and mobile users can collaborate in geographically widespread regions, using cognitive radio and geographic databases of users. This White Space (WS) standard is well described in the literature~\cite{white, book} and is not the major topic of this short paper.

We wish to extend the idea of the WS concept to include the idea of EM sensors (such as Radar) adopting this approach to spectrum sharing, providing a quantum leap in access to spectrum. We postulate that networks of sensors, using the tools developed by the WS community, can replace and enhance our present set of EM sensors.

We first define what Networks of Sensors entail (with some history), and then go on to define, based on a Taxonomy of Symbiosis defined by de Bary\cite{symb}, how these sensors and other users (especially communications) can co-exist. This new taxonomy is important for understanding, and should replace somewhat outdated terminologies from the radar world.

\section{Networks of Sensors}

Here we discuss the migration from what might be perceived as a single or limited sensor function such as a radar, to the concept of a network of sensors using electromagnetic waves to detect, classify, identify objects in a volume of interest. How this is achieved, is, of course, highly variable in terms of architecture, operating frequency band, bandwidth and measurement method.

\subsection{Early networks of sensors}

Aircraft sensors using EM waves were developed by many nations in the Second World War, but the term, ``Radar'' (Radio Detection and Ranging) evolved in the USA and has been adopted universally.  Radar represents a class of sensor that emitted an electromagnetic (EM) wave and by measuring time delay of the echo from a target, distance to target was inferred. Rapidly the sensor evolved to measure Doppler (radial velocity), and, using directional antenna technology, a measure of bearing angle or even, elevation angle became routine.

H\"ulsmeyer patented a sensor modelled on Marconi's communications technology that could determine the presence of a reflecting object. It did not have the capability inferred in the term, ``Radar'' of measuring range, and assumed that only objects in the common volume of the transmitter and receiver were important.

\subsection{Taxonomy for Networks of Sensors}

We have chosen the term, ``sensor'' for the technology that will be revealed in this short paper, but more specifically, a class of sensors that uses EM waves, but in a symbiosis with other uses of the same EM waves. To understand the subtlety of of the term ```symbiosis'', we have adopted a classification used in Biology, when a similar controversy existed in describing the way certain organisms (plants and animals) coexisted. Anton de Bary in a paper~\cite{symb} defined the interaction of mutually dependent organisms as ``symbiosis'', with three subdivisions i.e.:

\subsubsection{Symbiosis}

\begin{figure}[htbp]
\begin{center}
\centering{\includegraphics[scale=.4]{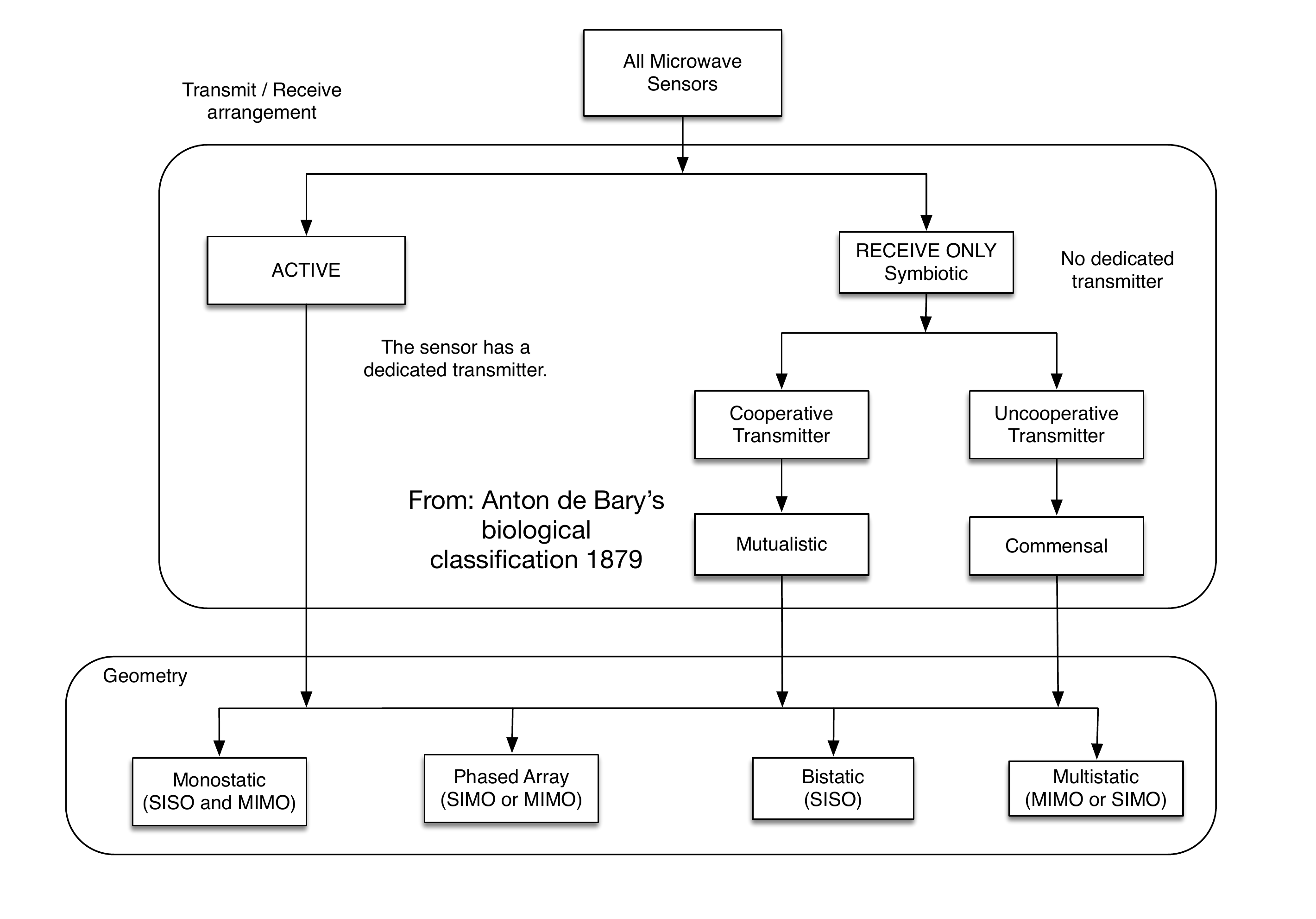}}
\caption{A Taxonomy for a network of sensors showing possible collaboration with other users of the EM spectrum.}
\label{fig:tax}
\end{center}
\end{figure}

De Bary saw three types of relationship existing between organisms, which we can explore, adding the context for sensor networks. A family tree of EM sensors fulfilling Radar type sensing functions is shown in Figure~\ref{fig:tax}.
\begin{description}

\item[Parasitic Systems]

In nature, parasitism leads to a degradation of one organism due to the parasite harvesting resources of the host, often leading to the death of the host. There does not seem to an analogue in the EM sensor network situation. 

\item[Commensal Systems]

Commensalism between organisms implies that neither system degrades the other, but in the sensor / communications symbiosis, we note that, for example, the sensor might not exist if it were not for the commensal partner. The simplest example is the case of a sensor system that utilises the FM Broadcast Band emissions to track aircraft~\cite{inggs1, inggs2}. In particular, Inggs et al.~\cite{inggs1} describe such systems and how they are implemented in some detail. 

\item[Mutualistic Systems]

Here, two organisms collaborate to mutual benefit i.e. the two functions (say telecommunications and sensing) are designed together, and compromises are made to ensure efficient operation of both functions. The White Space / Telecommunication symbiosis , is described in a patent~\cite{patent}.
\end{description}

Thus, in the WS Band, we have shown that the EM sensing function of tracking aircraft for air traffic management (ATM) can be carried out using FM Band broadcasts. Since this is a Commensal sensor, only the presence of stable broadcast transmitters is required. We can envisage more complex scenarios, where a Mutualistic relationship exists i.e. the sensor is in fact a mobile user of a WS network, and uses the network itself to share detections of targets between geographically dispersed nodes, thereby solving may of the tracking problems. These are discussed in other papers, but the field is wide open to exploitation.

At the bottom of the taxonomy chart, we refer to the geometric configuration of the sensor network i.e. the spatial distribution of the transmitters and receivers. We do not elaborate further on this aspect in the paper.

\section{Conclusions}
We have demonstrated that EM sensors, described by a taxonomy based on the Biological concepts of Symbiosis will replace the current, ``hard wired'' sensors, some with heritage going back to before the 1939-1945 World War, and an unwieldy terminology.

The WS collaboration has shown the way. The hope is that sensor designers will see the advantages of geographically and frequency distributed sensors that can seamlessly collaborate with other spectrum users, leading to a new world where spectrum congestion can be eased. It is not too far far fetched to maybe predict:

\begin{description}
\item[Radiometers ] A fleet of L Band radiometers for soil moisture observation from Space, collaborating with L Band Communications and Radar users as the satellite passes over, to ensure a clean spectrum for the soil moisture measurments.
\item[Air Traffic Radar] A new network of radars using a single frequency, but scheduling transmissions with nearby system for mutual convenience.
\item[Monitoring Infirm] A sensor network in a house to monitor the movement of the infirm, checking for possible falls, or, unexpected inactivity~\cite{chetty:12,chen:16, tan:16, tan:15}.
\item[Imaging Radar] A fleet of satellite, airborne and ground based radars that illuminated the earth in a coordinated way, extracting different imaging information, based on user needs.
\end{description}

We hope that papers such as this will start the EM sensor community to start thinking laterally about collaboration, and the financial incentives offered by sharing. Clearly, no one wishes for performance standards to be degraded, but this should not be necessary.

\printbibliography

\end{document}